\documentclass[11pt]{article}

\usepackage{latexsym}
\usepackage{amsmath,amssymb,amsfonts}

\usepackage{orcidlink}

\def\mymedskip{\vskip\medskipamount}
\def\mymedbreak{\par \ifdim\lastskip<\medskipamount
  \removelastskip \penalty-100 \mymedskip \fi}
\def\myaftermedspace{\par \ifdim\lastskip<\medskipamount
  \removelastskip \penalty55\mymedskip\fi}
\newcommand{\eop}{{\unskip\nobreak\hfil\penalty50
          \hskip2em\hbox{}\nobreak\hfil$\Box$
          \parfillskip=0pt \finalhyphendemerits=0 \par}}
\newenvironment{proof}%
{\mymedbreak{\noindent\bf Proof.\enspace}}{\eop\myaftermedspace}
\newenvironment{proofn}[1]%
{\mymedbreak{\noindent\bf Proof #1.\enspace}}{\eop\myaftermedspace}
{\mymedbreak{\noindent\bf Proof of Theorem~\ref{#1}:\enspace}}{\eop\myaftermedspace}
{\mymedbreak\noindent{\bf Remark:}%
\enspace\rm}%
{\myaftermedspace}
\newtheorem{teor}{Theorem}[section]
\newtheorem{defi}[teor]{Definition}
\newtheorem{fact}[teor]{Fact}
\newtheorem{problem}{Problem}
\newtheorem{exercise}{Exercise}
\newtheorem{examp}[teor]{Example}
\newtheorem{lem}[teor]{Lemma}
\newtheorem{cor}[teor]{Corollary}
\newtheorem{con}[teor]{Conjecture}
\newtheorem{prop}[teor]{Proposition}
\newtheorem{rem}[teor]{Remark}
\newtheorem{cons}[teor]{Construction}
\newcommand{\beq}{\begin{equation}}
\newcommand{\eeq}{\end{equation}}
\newcommand{\beql}[1]{\begin{equation} \label{#1}}
\newcommand{\eeql}{\end{equation}}
\newcommand{\beqa}{\begin{eqnarray*}}
\newcommand{\eeqa}{\end{eqnarray*}}
\newcommand{\beqal}[1]{\begin{eqnarray} \label{#1}}
\newcommand{\eeqal}{\end{eqnarray}}
\newcommand{\beqan}{\begin{eqnarray}}
\newcommand{\eeqan}{\end{eqnarray}}
\newcommand{\bpf}{\begin{proof}}
\newcommand{\epf}{\end{proof}}
\newcommand{\bpfn}[1]{\begin{proofn}{#1}}
\newcommand{\epfn}{\end{proofn}}
\newcommand{\ben}{\begin{enumerate}}
\newcommand{\een}{\end{enumerate}}
\newcommand{\bit}{\begin{itemize}}
\newcommand{\eit}{\end{itemize}}

\newcommand{\bab}{\begin{abstract}}
\newcommand{\eab}{\end{abstract}}
\newcommand{\bke}{\begin{keywords}}
\newcommand{\eke}{\end{keywords}}

\newcommand{\btm}[1]{\begin{teor} \label{#1}}
\newcommand{\etm}{\end{teor}}
\newcommand{\btmn}[2]{\begin{teor}[#1] \label{#2}}
\newcommand{\etmn}{\end{teor}}
\newcommand{\ble}[1]{\begin{lem} \label{#1}}
\newcommand{\ele}{\end{lem}}
\newcommand{\bLe}[1]{\begin{Lemma} \label{#1}}
\newcommand{\eLe}{\end{Lemma}}
\newcommand{\blen}[2]{\begin{lem}[#1] \label{#2}}
\newcommand{\elen}{\end{lem}}
\newcommand{\bpn}[1]{\begin{prop} \label{#1}}
\newcommand{\epn}{\end{prop}}
\newcommand{\bex}[1]{\begin{examp} \label{#1}}
\newcommand{\eex}{\eop\end{examp}}
\newcommand{\bde}[1]{\begin{defi} \label{#1}}
\newcommand{\ede}{\end{defi}}
\newcommand{\bco}[1]{\begin{cor} \label{#1}}
\newcommand{\eco}{\end{cor}}
\newcommand{\bcorn}[2]{\begin{cor}[#1] \label{#1}}
\newcommand{\ecorn}{\end{cor}}
\newcommand{\bcon}[1]{\begin{con} \label{#1}}
\newcommand{\econ}{\end{con}}
\newcommand{\bfa}[1]{\begin{fact} \label{#1}}
\newcommand{\efa}{\end{fact}}
\newcommand{\bpr}[1]{\begin{problem} \label{#1}}
\newcommand{\epr}{\end{problem}}
\newcommand{\bprnn}[1]{\begin{problemnn} \label{#1}}
\newcommand{\eprnn}{\end{problemnn}}
\newcommand{\bprn}[2]{\begin{problem}[#1] \label{#2}}
\newcommand{\eprn}{\end{problem}}
\newcommand{\bexer}[1]{\begin{exercise} \label{#1}}
\newcommand{\eexer}{\end{exercise}}
\newcommand{\bre}[1]{\begin{rem} \label{#1}}
\newcommand{\ere}{\end{rem}}
\newcommand{\bcons}[1]{\begin{cons} \label{#1}}
\newcommand{\econs}{\end{cons}}
\newcommand{\Tm}[1]{Theorem~\protect\ref{#1}}
\newcommand{\Le}[1]{Lemma~\protect\ref{#1}}

\newcommand{\Sec}[1]{Section~\protect\ref{#1}}

\newcommand{\bbF}{{\mathbb{F}}}
\newcommand{\F}{{\mathbb{F}}}
\newcommand{\bbN}{{\mathbb{N}}}

\newcommand{\bbZ}{{\mathbb{Z}}}

\newcommand{\cL}{{\cal L}}

\newcommand{\cU}{{\cal U}}

\usepackage[a4paper,margin=27mm]{geometry}
\hypersetup{hidelinks}
\DeclareMathOperator{\ord}{ord}
\DeclareMathOperator{\PAut}{PAut}
\DeclareMathOperator{\GL}{GL}
\DeclareMathOperator{\lcm}{lcm}
\DeclareMathOperator{\Tr}{Tr}

\newcommand{\NSset}{\mathcal N}
\newcommand{\CN}{{N}}
\newcommand{\ESset}{\mathcal E}
\newcommand{\PrimeSet}{\mathcal P}
\title{A density theorem for non-standard irreducible cyclic codes}
\author{\normalsize Tao Feng, Henk D. L. Hollmann\thanks{Corresponding author.}, Weicong Li, Qing Xiang}
\date{First draft, 24 October 2026}
\begin{document}
\maketitle
\begin{abstract}
We prove a fixed-characteristic density form of the Berger-Charpin conjecture
for non-degenerate irreducible cyclic codes: for each prime $p$, the
lengths admitting such a non-standard code over some finite field of
characteristic~$p$ have natural density zero. Thus, for almost every length coprime to $p$, every such
code over every finite field of characteristic $p$ has only the
permutation automorphisms generated by the cyclic shift and the
Frobenius map. The proof uses the
Feng--Hollmann--Li--Xiang classification for non-standard irreducible cyclic codes, a normal form for the deriving constructions in that paper, and a theorem by Wiertelak on multiplicative orders, in the form recalled by Moree.
\end{abstract}
\begin{center}
\end{center}

\section{Introduction}
A $q$-ary linear code $C$ of length $n$ is an $\F_q$-linear subspace
of $\F_q^n$, where $q$ is a prime power. The permutations of the
coordinate positions that preserve $C$ form its \emph{permutation
automorphism group}, denoted by $\PAut(C)$. We label the positions
by $\bbZ_n$ and call the code cyclic if this group contains the shift
$\sigma:i\mapsto i+1$.

We assume throughout that $\gcd(n,q)=1$ and consider non-degenerate
irreducible cyclic codes. Such a code is called \emph{standard} if
its permutation automorphism group is generated by the shift and
the Frobenius map $\psi:i\mapsto qi$, and \emph{non-standard}
otherwise. Standardness depends only on the length and alphabet \cite[Proposition 5.19]{HollmannI},
so we may speak of a standard or non-standard pair $(n,q)$. The necessary
background is recalled in Section~\ref{sec:pairs}.

Charpin~\cite[Section 3.5]{Charpin}, motivated by the work of Berger
and Charpin~\cite{BergerCharpin}, conjectured that cyclic codes
generally have small permutation automorphism groups, generated
by the shift and a Frobenius mapping depending on the alphabet.
For general cyclic codes, further affine permutations may arise
from symmetries of the defining zeros. It is therefore natural
to expect that non-degenerate cyclic codes usually have only
affine permutation automorphisms, called standard automorphisms
in~\cite[Theorem 5.28, Definition 5.29]{HollmannI}.
For non-degenerate irreducible cyclic codes, these are precisely
the permutations generated by the shift and the Frobenius
multiplier~\cite[Corollary 5.17]{HollmannI}.

The classification of Feng, Hollmann, Li and Xiang~\cite{FHLX}
makes it possible to give this expectation a density interpretation
for non-degenerate irreducible cyclic codes. Their classification lists
five families of non-standard pairs and four constructions from which
all other non-standard pairs arise. The numerical evidence
in~\cite[Section 6.3]{FHLX} suggests that the proportion of exceptional lengths tends to zero for each fixed alphabet.

The non-degeneracy assumption is essential. As shown
in~\cite[Lemma 2.4]{FHLX}, every nonzero degenerate cyclic code
has permutation automorphisms outside the affine group, apart from
four exceptions of length at most $4$. In particular, for every
$n\ge4$ coprime to $q$, the repetition code is irreducible,
degenerate and non-standard. 
Thus, if degenerate codes are allowed, the
lengths admitting a non-standard irreducible cyclic code have relative density 1 among the integers coprime to~$q$.

For a prime power $q=p^r$, let $\NSset_q$ be the set of positive
integers $n$ coprime to $p$ for which $(n,q)$ is non-standard, and define
\begin{equation}\label{eq:ratio}
 \CN_q(X)=\frac{\#\{n\le X:n\in\NSset_q\}}
                  {\#\{n\le X:\gcd(n,p)=1\}}.
\end{equation}
We prove that $\CN_q(X)\to0$. In fact, the exceptional lengths still
have density zero when all alphabets of a fixed characteristic are
considered together.

\btm{thm:main}
For every prime $p$, $\bigcup_{r\ge1}\NSset_{p^r}=\NSset_p$ and
\begin{equation}\label{eq:main-density}
 \#\{n\le X:n\in\NSset_p\}=o(X)
 \qquad\text{as }X\longrightarrow\infty.
\end{equation}
Thus, for almost every length coprime to $p$, every non-degenerate
irreducible cyclic code of that length over every finite field of characteristic~$p$ is standard.
\etm
Here ``almost every'' is relative to the positive integers coprime
to $p$. For a fixed alphabet, lengths also index the
permutation-equivalence classes of non-degenerate irreducible
cyclic codes.

Theorem~\ref{thm:main} establishes Charpin's expectation for
non-degenerate irreducible cyclic codes in a precise density
sense, simultaneously over all finite fields of a fixed
characteristic. In~\cite{FHLX}, the underlying conjecture is
called the Berger--Charpin conjecture.
\bco{cor:ratios}
For each prime $p$, $\sup_{r\ge1}\CN_{p^r}(X)=\CN_p(X)\to0$ as
$X\to\infty$.
\eco
\bpf
The union in Theorem~\ref{thm:main} equals $\NSset_p$, so
$\CN_{p^r}(X)\le \CN_p(X)$ for every $r\ge1$, with equality at $r=1$.
The limit follows because the common denominator is
$(1-1/p)X+O(1)$.
\epf

A difficulty is that the classification allows arbitrarily
long sequences of constructions to obtain new non-standard pairs from a given one (all operating within the same field characteristic). In this paper, we first show that all these constructions can be described in terms of three basic constructions called {\em lifting\/} (moving to an extension field), {\em extension\/} (multiplying the length by a factor), and {\em descent\/} (moving to a subfield), see Section~\ref{sec:derived}. Then we show that every such sequence of constructions reduces to one lift, one extension and one descent, see also~\cite{HollmannII}. 

For a fixed initial pair $(n,q)$ of degree greater than one,
every derived length is $nu$, where $u$ avoids a fixed set
of primes whose reciprocals have divergent sum, by a theorem
of Wiertelak~\cite{Wiertelak,Moree}.
A sieve argument then demonstrates density zero. To pass from one initial pair to an entire family, we use summable tails for the primitive and ovoid families and a common sieve for the duals of repetition codes.
The equally-spaced family requires a separate argument: it is closed
under the constructions, and its lengths can be counted using a
prime at which the multiplicative order increases.

The paper is organized as follows. 
In Section~\ref{sec:pairs}, we recall the classification of
non-standard pairs. Section~\ref{sec:derived} describes the
derived pairs and proves the normal form, and
Section~\ref{sec:closed} establishes closure of the equally-spaced
family. After collecting the number-theoretic tools in
Section~\ref{sec:density-lemmas}, we prove our main results in~\Sec{sec:main}. The density results
for the five families are proved in Sections~\ref{sec:four}
and~\ref{sec:es-density}, and we complete the proof of Theorem~\ref{thm:main} in~Section~\ref{sec:proof}. In Section~\ref{sec:conclusions} we present some conclusions.


While completing this paper, we learned that Wu and
Xiang~\cite{WuXiang} have recently obtained density results
concerning permutation automorphisms common to all cyclic
codes of a given length over a fixed field. Their proof uses
the Wiertelak--Moree theorem and related sieve and tail
arguments. We study the full permutation automorphism group
of an individual non-degenerate irreducible cyclic code:
a permutation preserving such a code need not preserve all
cyclic codes of the same length. Our proof combines these
number-theoretic tools with the classification and the normal
form for derived pairs, using Lemma~\ref{lem:tails} to handle
entire families of derived lengths. Moreover, our density
result holds simultaneously over all finite fields of a fixed
characteristic.

\section{Preliminaries}\label{sec:pairs}

Let $p$ be a prime, let $q$ be a power of $p$, and let $\F_q$ be the
finite field of size $q$. We use $\bbN$ for the positive integers
and take all field extensions inside a fixed algebraic closure.

Identifying $(c_0,\ldots,c_{n-1})$ with $\sum_i c_i x^i$, a cyclic
code becomes an ideal of $\F_q[x]/(x^n-1)$. It is generated by a
unique monic divisor $g(x)$ of $x^n-1$. Its \emph{check polynomial}
is $h(x)=(x^n-1)/g(x)$, and we write $C_{n,h,q}$ for the code.
The order of $h(x)$ is the least positive integer $k$ for which
$h(x)\mid x^k-1$. The code is non-degenerate when this order is $n$,
and irreducible when $h(x)$ is irreducible, or equivalently when the
code is a minimal nonzero ideal. See~\cite{vanLint,MacWilliamsSloane}
for background, and~\cite{FiniteFields} for finite-field notation
and facts.

For $n\ge1$ coprime to $q$, let $m=\ord_n(q)$, the least positive
integer for which $n\mid q^m-1$, with $\ord_1(q)=1$.
We commonly refer to $m$ as the \emph{degree} of the pair $(n,q)$.
Let $\cU_{n,q}$ be the subgroup of order $n$ in $\F_{q^m}^*$,
and choose a generator $\xi$. Its powers span $\F_{q^m}$ over
$\F_q$, since $\F_q(\xi)=\F_{q^m}$. If $h(x)$ is the minimal
polynomial of $\xi^{-1}$ over $\F_q$, then $h(x)$ has degree $m$
and order $n$. The code $C_{n,h,q}$ has dimension $m$ and the
trace representation
\begin{equation}\label{eq:trace}
 C_{n,h,q}=
 \{(\Tr(b\xi^i))_{i=0}^{n-1}:
       b\in\F_{q^m}\},
\end{equation}
where $\Tr$ is the trace from $\F_{q^m}$ to $\F_q$;
see~\cite[Section 3.4.3]{Charpin} and
\cite[Theorem 5.25]{PlessHuffmanBrualdi}. Every non-degenerate
irreducible cyclic code of length $n$ over $\F_q$ arises in this
way by choosing $\xi$ so that $\xi^{-1}$ is a root of the check polynomial, as in~\cite[Section 2.5]{FHLX}.

Replacing $\xi$ by another generator $\xi^u$ permutes the
coordinates in~\eqref{eq:trace} by a multiplier. Thus all these
codes are equivalent, as observed
in~\cite[Proposition 5.19]{HollmannI}. Coordinate multipliers
normalize $\langle\sigma,\psi\rangle$, so standardness is
preserved. It is also unchanged by the choice of finite-field
model, since field isomorphisms commute with the trace and
coordinate permutations.

For the permutation automorphism groups, define
\[
\begin{aligned}
 \cL(n,q)&=\{T\in\GL_{\F_q}(\F_{q^m}):
                T(\cU_{n,q})=\cU_{n,q}\},\\
 \cL_{\rm st}(n,q)
 &=\{x\mapsto c x^{q^j}:c\in\cU_{n,q},\ 0\le j<m\}.
\end{aligned}
\]
Hollmann~\cite[Corollary 5.14 and Theorem 5.15]{HollmannI} gives
the following identification.

\begin{prop}\label{prop:linear-model}
Let $C=C_{n,h,q}$ be as above. The affine permutations in
$\PAut(C)$ form the subgroup $\langle\sigma,\psi\rangle$,
of order $nm$. There is an isomorphism $\PAut(C)\cong\cL(n,q)$
taking this subgroup to $\cL_{\rm st}(n,q)$. In particular,
$(n,q)$ is non-standard if and only if $\cL(n,q)\ne\cL_{\rm st}(n,q)$.
\end{prop}
In the terminology of non-standard linear recurring sequence
subgroups, $(n,q)$ is non-standard precisely when $\cU_{n,q}$
is a non-standard subgroup over $\F_q$. This connects the present
problem with the work of Brison and Nogueira~\cite{BrisonNogueira};
see~\cite{HollmannI} for the correspondence with cyclic codes.

If $m=1$, every invertible linear map is a scalar multiplication,
so every pair of degree one is standard.

The classification starts from the following five
families~\cite[Section 3]{FHLX}, which may overlap.
\begin{enumerate}
\item[\rm (i)] \emph{Duals of repetition codes.} The pairs $(n,p)$,
where $p$ and $n$ are primes, $n\ge5$, and $\ord_n(p)=n-1$.
\item[\rm (ii)] \emph{Duals of primitive BCH codes of designed
distance $2$.} The pairs $(q^m-1,q)$, where $q$ is a prime power,
$m\ge2$, and $(q,m)\ne(2,2)$.
\item[\rm (iii)] \emph{Extension ovoid codes.} The pairs
$((q-1)(q^2+1),q)$, where $q$ is a prime power. Their degree is $4$.
\item[\rm (iv)] \emph{Duals of Golay codes.} The pairs $(23,2)$ and
$(11,3)$, of degrees $11$ and $5$, respectively.
\item[\rm (v)] \emph{Equally-spaced check polynomials.} The pairs
$(n,q)$ with $n\ge6$ for which there exist integers $d,s\ge2$ such that
\begin{equation}\label{eq:es}
 n=ds,\qquad \ord_n(q)=s\ord_d(q).
\end{equation}
\end{enumerate}

For family (v), let $a$ be a root of the check polynomial~$h(x)$. Then~$a$ has order~$n$. Let~$h_0(x)$ be the minimal polynomial of~$a^s$ over~$\bbF_q$. Then $a^s$ has order~$d$, and $h_0(x^s)$ has root $a$ and degree
$s\ord_d(q)=\ord_n(q)$. Hence the check polynomial is
$h(x)=h_0(x^s)$. If $g_0(x)=(x^d-1)/h_0(x)$, the generator
polynomial is $g_0(x^s)$. Grouping coordinates according to their
residues modulo $s$ gives a direct sum of $s$~copies of
$C_{d,h_0,q}$; see~\cite[Example 3.5 and Section 4]{FHLX} for
their permutation automorphism groups.
The length-$4$ case satisfying~\eqref{eq:es} is standard and is
excluded from the initial family.

In what follows, we will refer to these five families as the {dual-repetition\/} family, the {\em primitive\/} family, the {\em ovoid\/} family, the {\em Golay\/} family, and the {\em equally-spaced\/} family.

\btmn{Feng--Hollmann--Li--Xiang}{thm:classification}
Every non-standard pair can be obtained from one of the five families
above by a finite sequence of the following constructions, in which
the characteristic remains fixed:
\begin{enumerate}
\item[\rm (i)] extension $(n,q)\mapsto(nu,q)$, where $u\ge1$ and
$u\mid(q-1)/\gcd(n,q-1)$;
\item[\rm (ii)] lifting $(n,q)\mapsto(n,q^t)$, where $t\ge1$ and
$\gcd(t,\ord_n(q))=1$;
\item[\rm (iii)] special descent $(n,q^r)\mapsto(n,q)$, where
$n\ge4$ and $r>1$ divides $\ord_n(q)$;
\item[\rm (iv)] the tensor-product construction
$(n,q)\mapsto(\lcm(n,s),q)$, where $s>1$, $\gcd(s,q)=1$,
$t=\ord_s(q)>1$ and $\gcd(t,\ord_n(q))=1$.
\end{enumerate}
The initial pairs are non-standard, and each construction preserves
non-standardness.
\etmn

This is~\cite[Theorem 1.1 and Section 3]{FHLX} in the language of
pairs. Its proof uses the classification of finite simple groups.

\section{Derived pairs and a normal form}\label{sec:derived}
Our aim is to describe the lengths of all pairs obtainable from a given
pair $(n,q)$ by the constructions in Section~\ref{sec:pairs}.
We first introduce general descent and express the tensor construction
as a lift, an extension and a descent. We then prove that, in fact, every sequence
of these three operations has this form. The resulting description of
the lengths is given in Corollary~\ref{cor:derived-lengths}.

Throughout this section, pairs $(n,q)$ have $n\ge1$, $q$ a prime power
and $\gcd(n,q)=1$. We retain the lifting and extension operations of
Section~\ref{sec:pairs}; we explicitly allow identity operations for each of them. The general form of descent
was noted in~\cite[Remark 3.12]{FHLX}. By a \emph{descent} we mean
\[
 (n,q)\longmapsto(n,q_0),\qquad q=q_0^r,
\]
where $q_0$ is a prime power and $r\ge1$ is an integer. Again, we explicitly allow the identity operation (i.e., $q_0=q$ and $r=1$). There is no
condition on $r$ relative to the degree. In particular, special descent
is an instance of descent. 
\begin{defi}\label{def:derived}
A pair $(n',q')$ is called a \emph{derived pair} of $(n,q)$
if it can be obtained from $(n,q)$ by a finite (possibly empty) sequence of
lifts, extensions and descents.
\end{defi}

\bpn{thm:operations}
Lifting, extension and descent preserve non-standardness. Consequently, every pair derived from a non-standard pair
is itself non-standard.
\epn
\bpf
For lifting and extension this is recalled in
Theorem~\ref{thm:classification}. For descent, this is proved in~\cite{HollmannII}; for completeness we provide a short proof here.
Write $q=q_0^r$, and write
\[
 m_0=\ord_n(q_0),\qquad d=\gcd(m_0,r),\qquad
 m=\ord_n(q)=m_0/d.
\]
The subgroup $H$ of order~$n$ is the same in~$\bbF_{q_0^{m_0}}$ and~$\bbF_{q^m}$:
$H=\cU_{n,q_0}=\cU_{n,q}$. It spans $\F_{q_0^{m_0}}$ over $\F_{q_0}$ and
$\bbF_{q^m}$ over $\F_q$. Consequently restriction gives an injective homomorphism
\[
 \rho:\cL(n,q)\longrightarrow\cL(n,q_0).
\]
Indeed, a map preserving $H$ preserves its $\F_{q_0}$-span $\F_{q_0^{m_0}}$,
and its action on $H$ determines the map on $\F_{q^m}$.

Standard maps restrict to standard maps. Conversely, suppose that
$T\in\cL(n,q)$ restricts to $x\mapsto c x^{q_0^j}$ on $\F_{q_0^{m_0}}$,
where $c\in H$. Replacing $T$ by $c^{-1}T$, we may assume $c=1$.
This normalized map fixes $\F_q$ pointwise. Since
\[
 \F_{q_0^{m_0}}\cap\F_q=\F_{q_0^d},
\]
we obtain $d\mid j$. The congruence $rv\equiv j\pmod{m_0}$
therefore has a solution $v$. The standard map $x\mapsto x^{q^v}$
on $\F_{q^m}$ has the same restriction to $\F_{q_0^{m_0}}$ as $T$. Injectivity of $\rho$
makes these maps equal. Thus a non-standard map remains non-standard
under restriction, proving the assertion by
Proposition~\ref{prop:linear-model}.
\epf

\bco{cor:prime-reduction}
For each prime $p$,
\[
 \bigcup_{r\ge1}\NSset_{p^r}=\NSset_p.
\]
\eco
\bpf
Descent shows $\NSset_{p^r}\subseteq\NSset_p$ for every $r>1$.
\epf

For a pair $(n,q)$ of degree $m=\ord_n(q)$, we call an integer
$u\ge1$ an \emph{extension factor} if
$u\mid(q-1)/\gcd(n,q-1)$, and an integer $t\ge1$ a
\emph{lift exponent} if $\gcd(t,m)=1$.
Thus the operation 
$(n,q)\mapsto(nu,q)$ is an extension precisely when $u$ is an extension factor for~$(n,q)$, and the operation $(n,q)\mapsto(n,q^t)$ is a lift precisely when~$t$ is a lift exponent for~$(n,q)$.

\ble{lem:arith}
Let $(n,q)$ be a pair of degree $m=\ord_n(q)$.
Then the following hold.\\
(i) The pair $(n',q)$ is an extension of $(n,q)$ if and only if $n\mid n'\mid\lcm(n,q-1)$.\\
(ii)
If $(n',q)$ is an extension of $(n,q)$, then $\ord_{n'}(q)=\ord_n(q)$ and $\lcm(n',q-1)=\lcm(n,q-1)$.\\
(iii)
For every integer $t\ge1$, 
an  extension factor for~$(n,q)$ is also an extension factor for~$(n,q^t)$. \\
(iv)
For every integer $t\ge1$, we have $\ord_n(q^t)=m/\gcd(m,t)$.  
\ele
\bpf
(i) 
Write $n'=nu$ and use $\lcm(n,q-1)/n=(q-1)/\gcd(n,q-1)$.\\
(ii) By (i), $n\mid n'\mid\lcm(n,q-1)\mid q^m-1$, so $\ord_{n'}(q)=m$. The equality of least common multiples also follows from (i).\\
(iii) Since $q-1\mid q^t-1$, we have $\lcm(n,q-1)/n
\mid\lcm(n,q^t-1)/n$.
Now apply the identity in the proof of (i).\\
(iv) The order of $q^t$ modulo $n$ is the least positive
integer $j$ for which $m\mid tj$, which equals  $m/\gcd(m,t)$.
\epf

\bpn{prop:tensor}
Every construction in Theorem~\ref{thm:classification} can be realized
by a lift, an extension and a descent, allowing identity operations.
Consequently, the non-standard pairs are exactly the pairs derived
from the five initial families.
\epn
\bpf
Only the tensor construction requires consideration. Let
$m=\ord_n(q)$, $t=\ord_s(q)$ and $n'=\lcm(n,s)$, so that
$\gcd(m,t)=1$. Use the sequence
\[
 (n,q)\rightarrow(n,q^t)\rightarrow(n',q^t)
                   \xrightarrow{\mathrm{dec}}(n',q).
\]
Since $\gcd(t,m)=1$, the first step is a lift. Since $s\mid q^t-1$, we have $\lcm(n,s)\mid \lcm(n,q^t-1)$, hence the second step is an extension by~\Le{lem:arith}, part~(i). The third is a descent by definition.
The final claim follows from the classification in~\Tm{thm:classification} and
Proposition~\ref{thm:operations}.
\epf

\btm{thm:normal}
Every finite sequence of lifts, extensions and descents can be replaced
by one lift, followed by one extension and one descent, with the same
initial and final pairs.
As a consequence, if $q=p^e$ and $m=\ord_n(q)$, then $(n',p^{e'})$
is derived from $(n,q)$ if and only if $n'=nu$ for integers $u,t\ge1$
satisfying
\begin{equation}\label{eq:normal}
 \gcd(t,m)=1,\qquad
 u\mid\frac{q^t-1}{\gcd(n,q^t-1)},\qquad e'\mid et.
\end{equation}
\etm
\bpf
By \Le{lem:arith}, part~(iii), a descent-extension sequence
$
 (n,q^r)\xrightarrow{\mathrm{desc}}(n,q)\xrightarrow{\mathrm{ext}}(nu,q)
$
 can be replaced by the extension-descent sequence
 $
 (n,q^r)\xrightarrow{\mathrm{ext}}(nu,q^{r})\xrightarrow{\mathrm{desc}}(nu,q)
$.
Similarly, by~\Le{lem:arith}, parts~(ii) and~(iii), an extension-lift sequence
$
 (n,q)\xrightarrow{\mathrm{ext}}(nu,q)\xrightarrow{\mathrm{lift}}(nu,q^t)
$
 can be replaced by the lift-extension sequence
$
 (n,q)\xrightarrow{\mathrm{lift}}(n,q^{t})\xrightarrow{\mathrm{ext}}(nu,q^t)
$.
Finally, by~\Le{lem:arith}, part~(iv), a descent-lift sequence 
$
 (n,q^r)\xrightarrow{\mathrm{desc}}(n,q)\xrightarrow{\mathrm{lift}}(n,q^t)
$
can be replaced by the lift-descent sequence
$
 (n,q^r)\xrightarrow{\mathrm{lift}}(n,q^{rt})\xrightarrow{\mathrm{desc}}(n,q^t)
 $.

These replacements move all lifts first,  followed by all extensions, and all descents last. 

By~\Le{lem:arith}, part (iv), consecutive lifts combine into one lift;  consecutive
extensions combine into one by~\Le{lem:arith}, parts~(i) and~(ii), and consecutive
descents obviously combine into one. 

The conditions in~\eqref{eq:normal} are precisely those
for the sequence
$(n,p^e)\xrightarrow{\mathrm{lift}}(n,p^{et})\xrightarrow{\mathrm{ext}}(nu,p^{et})\xrightarrow{\mathrm{desc}}(nu,p^{e'})$.
\epf

For a pair $(n,q)$ of degree $m=\ord_n(q)$, write
\beql{LEUn}
U_{n,q}:=\left\{u\in\bbN:\exists\,t\in\bbN,\quad
\gcd(t,m)=1,\quad
u\mid\frac{q^t-1}{\gcd(n,q^t-1)}\right\}.
\eeql
When $q$ is clear from the context, we simply write $U_n$ instead of~$U_{n,q}$.

\bco{cor:derived-lengths}
Fix a pair $(n,q)$ of characteristic~$p$, and let $m=\ord_n(q)$. The lengths of the pairs derived
from $(n,q)$ are precisely the integers in~$nU_n$.
Each such length occurs in a derived pair both over~$\bbF_q$ and over~$\bbF_p$.
\eco
\bpf
The description of the lengths follows from Theorem~\ref{thm:normal}.
For the last claim, take $e'=e$ and $e'=1$, respectively, in~\eqref{eq:normal}.
\epf

\section{The equally-spaced family is closed}\label{sec:closed}
We show that the equally-spaced family is closed under the
operations of lifting, extension, and descent discussed  in~Section~\ref{sec:derived}. We use the notation $n=ds$
of~\eqref{eq:es} and first record an order bound, which is also needed in~\Sec{sec:es-density}.

\ble{lem:order-bound}
If $d\mid n$ and $\gcd(n,q)=1$, then $\ord_d(q)\mid\ord_n(q)$ and
\[
 \frac{\ord_n(q)}{\ord_d(q)}\le\frac nd.
\]
\ele
\bpf
Let $m=\ord_n(q)$ and $k=\ord_d(q)$.
Since $d\mid n\mid q^m-1$, we have $k\mid m$.
Next, the numbers
\[
1,q^k,q^{2k},\ldots,q^{(m/k-1)k}
\]
are distinct modulo~$n$ and are all congruent to $1$ modulo~$d$.
Since there are exactly $n/d$ residue classes modulo $n$ congruent
to $1$ modulo $d$, we have $m/k\le n/d$.
\epf

\bpn{prop:closed}
Every pair derived from an equally-spaced non-standard pair is itself an
equally-spaced non-standard pair.
\epn
\bpf
By Proposition~\ref{thm:operations}, the three operations preserve
non-standardness. It remains to check the equally-spaced condition.
Consider an equally-spaced pair $(n,q)$. Suppose $n\ge6$, $n=ds$, with $d,s\ge2$, and
$\ord_n(q)=s\ord_d(q)$. Let $m=\ord_n(q)$ denote the degree of the pair $(n,q)$.

For a lift of exponent $t$ with $\gcd(t,m)=1$, we have  $\ord_n(q^t)=\ord_n(q)$ by~\Le{lem:arith}, part~(iv), so $(n,q^t)$ and $(n,q)$ have the same degree, and similarly $\ord_d(q^t)=\ord_d(q)$, so $(d,q^t)$ and $(d,q)$ have the same degree.  Hence the same $d,s$ work.

For an extension with extension factor $u$, we have $u\mid \lcm(n,q-1)/n=(q-1)/\gcd(n,q-1)$.
Since $d\mid n$, we also have
$u\mid(q-1)/\gcd(d,q-1)$.
Thus $(nu,q)$ and $(du,q)$ are extensions of $(n,q)$
and $(d,q)$, respectively. 
By Lemma~\ref{lem:arith}, part~(ii), both degrees are preserved, so
\[
nu=(du)s,\qquad \ord_{nu}(q)=s\ord_{du}(q).
\]

For a descent $(n,q^t)\mapsto(n,q)$, suppose the equally-spaced condition holds over $q^t$.
Let $m=\ord_n(q)$ and $m_d=\ord_d(q)$. By~\Le{lem:arith}, part~(iv), and since $m_d\mid m$, we have
\[
 s=\frac{m}{m_d}\frac{\gcd(m_d,t)}{\gcd(m,t)}
 \le\frac{m}{m_d}\le s,
\]
where the last inequality follows from Lemma~\ref{lem:order-bound}.
Thus $m=sm_d$, as required.

The length remains at least $6$ under all three operations.
The result follows by applying these observations successively.
\epf

\section{
\label{sec:density-lemmas}Wiertelak's theorem and some density results}
For $A\subseteq\bbN$, write $A(X)=|A\cap[1,X]|$. Its
\emph{natural density}, when the limit exists, and its
\emph{upper natural density} are respectively
\[
 d(A)=\lim_{X\to\infty}\frac{A(X)}X,
 \qquad \overline d(A)=\limsup_{X\to\infty}\frac{A(X)}X.
\]
Upper density zero implies natural density zero. To prove density zero,
we may replace a set by a larger set.

A set $\PrimeSet$ of primes is \emph{reciprocally divergent} if
$\sum_{r\in\PrimeSet}1/r=\infty$.
Its density \emph{within the primes} is
$\lim_{X\to\infty}\PrimeSet(X)/\pi(X)$, when this limit exists,
where $\pi(X)$ counts all primes of size at most $X$.
For integers $q>1$ and $h>1$, let
\beql{LEPqh}
 \PrimeSet(q,h)=\{r\text{ prime}:r\nmid q,\ h\mid\ord_r(q)\}.
\eeql

\btm{thm:wiertelak}
\textup{(Wiertelak~\cite{Wiertelak}; see Moree~\cite[Theorem 1 and
 the following paragraph]{Moree}.)}
For fixed integers $q>1$ and $h>1$, the set $\PrimeSet(q,h)$
has positive natural density within the primes. In particular,
$\PrimeSet(q,h)$ is reciprocally divergent.
\etm
%
Only the reciprocal divergence is needed here. Writing
$P(x)=|\PrimeSet(q,h)\cap[1,x]|$, the theorem and the prime
number theorem give $P(x)\sim\delta x/\log x$ for some
$\delta>0$. Hence
\[
\sum_{\substack{r\in\PrimeSet(q,h)\\x<r\le2x}}\frac1r
\ge\frac{P(2x)-P(x)}{2x}
\sim\frac{\delta}{2\log x}.
\]
Taking $x=2^j$ and summing over the disjoint intervals
$(2^j,2^{j+1}]$ proves divergence by comparison with
$\sum_j1/j$. (Here, any weaker elementary lower and upper bounds on $\pi(x)$ would suffice.)

We will also use the elementary bound
\begin{equation}\label{eq:primepower-sum}
 \sum_{r\text{ prime}}\sum_{a\ge2}\frac1{r^a}
 = \sum_{r\text{ prime}} \frac{1}{r(r-1)}
 =\sum_{r\text{ prime}}\left(\frac1{r-1}-\frac1r\right)
 \le\sum_{j=2}^{\infty}\left(\frac1{j-1}-\frac1j\right)=1.
\end{equation}
The inner sum is geometric, and the last sum telescopes.

\ble{lem:sieve}
Let $\PrimeSet$ be a reciprocally divergent set of primes.
The integers divisible by at most one of the primes in $\PrimeSet$ have density zero.
\ele
\bpf
For a finite subset $P\subseteq\PrimeSet$, divisibility by its primes
depends only on the residue class modulo $M:=\prod_{r\in P}r$. 
By the Chinese Remainder Theorem (CRT), the density of the set of integers avoiding all primes in $P$ is
\[
 D_P=M^{-1}\prod_{r\in P}(r-1)=\prod_{r\in P}(1-1/r).
\]
For a specific $r\in P$, the density of the set of integers that are divisible by $r$ and avoid all other primes in $P$ is
\[
 \frac1r\prod_{s\in P\setminus\{r\}}(1-1/s)=\frac1{r(1-1/r)}D_P=\frac{D_P}{r-1}.
\]
Since all relevant sets are disjoint, the density of the set of integers divisible by at most one prime in~$P$ is
\[
 D'_P=D_P\left(1+\sum_{r\in P}\frac1{r-1}\right).
\]
Let $S_P=\sum_{r\in P}1/r$. Since $1-x\le e^{-x}$,
we have $D_P\le e^{-S_P}$. Also, by~\eqref{eq:primepower-sum},
\[
 0\le\sum_{r\in P}\left(\frac1{r-1}-\frac1r\right)\le1.
\]
The two densities $D_P$ and $D'_P$ are therefore bounded
by $e^{-S_P}$ and $e^{-S_P}(2+S_P)$, respectively. These also bound
the upper densities of the corresponding sets for $\PrimeSet$.
Taking finite subsets with $S_P\to\infty$ proves our claim.
\epf

\ble{lem:tails}
Let $n_j$ be positive integers and let $A_j\subseteq n_j\bbN$ have
density zero for each $j\ge1$. If $\sum_j1/n_j<\infty$, then
$\bigcup_j A_j$ has density zero.
\ele
\bpf
This is a special case of Tannery's theorem (dominated convergence
for series; see, e.g., 
\cite[Chapter XVI (Tannery's Theorem)]{Fer38})
For completeness,
we give an elementary proof.
For each fixed $J$, since $A_j\subseteq n_j\bbN$,
\[
 \frac{|(\bigcup_{j\ge1}A_j)\cap[1,X]|}{X}
 \le \frac{|(\bigcup_{j\le J}A_j)\cap[1,X]|}{X}
       +\sum_{j>J}\frac1{n_j}.
\]
The first term tends to zero as $X\to\infty$, since a finite union
of density-zero sets has density zero. The tail tends to zero as
$J\to\infty$, since $\sum_j1/n_j$ converges. Taking these limits
in this order proves the result.
\epf

\section{Main results\label{sec:main}}
This section is devoted to the proof of \Tm{thm:main}. Throughout this section, $q$ will denote a fixed power of a prime~$p$. 
\subsection{Density zero for lengths derived from the first four families}\label{sec:four}
We prove in order that the lengths derived from the Golay families, the dual-repetition family, the primitive family, and the ovoid family
have density zero in each fixed characteristic. We first consider a fixed initial pair, then use a common sieve or summable tails to deal with the infinite families.
\bpn{prop:fixed}
Let $(n,q)$ be a fixed pair with $\gcd(n,q)=1$ and
$m=\ord_n(q)>1$. The set of lengths of all pairs derived
from $(n,q)$ has density zero.
\epn
\bpf
By Corollary~\ref{cor:derived-lengths}, the derived lengths constitute $nU_n$, where $U_n$ is as defined in~(\ref{LEUn}).
Let $u\in U_n$; then there is an integer $t\geq1$ such that $\gcd(m,t)=1$ and $u\mid q^t-1$, hence $\ord_u(q)\mid t$. Now recall the definition of $\PrimeSet(q,h)$ in~(\ref{LEPqh}).
If a prime $r\in\PrimeSet(q,m)$ divided $u$, then
\[
 m\mid\ord_r(q)\mid\ord_u(q)\mid t,
\]
contradicting $\gcd(t,m)=1$ since $m>1$. Thus $u$ has no prime divisor in
$\PrimeSet(q,m)$. This set of primes is reciprocally divergent by
Theorem~\ref{thm:wiertelak}, so $U_n$ has density zero by
Lemma~\ref{lem:sieve}. Since $n$ is fixed, it follows that $nU_n$ also has density zero.
\epf

\bco{prop:golay}
The lengths derived from the Golay pairs have density zero.
\eco
\bpf
Since $\ord_{23}(2)=11>1$ and $\ord_{11}(3)=5>1$, the result follows
from Proposition~\ref{prop:fixed}.
\epf

\bpn{prop:repetition}
For each fixed prime $p$, the lengths derived from all dual-repetition
pairs of characteristic $p$ have density zero.
\epn
\bpf
The initial pairs are $(n,p)$ with $n\ge5$ prime and
$\ord_n(p)=n-1$. By Corollary~\ref{cor:derived-lengths}, the
derived lengths from one such pair $(n,p)$ form $nU_n$, where $U_n$ is as defined in~(\ref{LEUn}). For $u\in U_n$, there is an integer
$t\ge1$ with $\gcd(t,n-1)=1$ and $u\mid p^t-1$, so
$\ord_u(p)\mid t$.
Since $n-1$ is even and $\gcd(t,n-1)=1$,
$t$ is odd. Thus $u$ has no prime divisor in $\PrimeSet(p,2)$,
the set of primes $s\nmid p$ for which $\ord_s(p)$ is even.
Consequently $nu$ has at most one prime divisor in the set $\PrimeSet(p,2)$, namely~$n$.
By Theorem~\ref{thm:wiertelak}, $\PrimeSet(p,2)$ is reciprocally
divergent; now Lemma~\ref{lem:sieve} shows density zero for the set of all derived lengths.
\epf

\bpn{prop:primitive-ovoid}
For each fixed prime $p$, the lengths derived from all primitive
and all ovoid pairs of characteristic $p$ have density zero.
\epn
\bpf
For the primitive family, let $n_{e,m}=p^{em}-1$, where $e\ge1$
and $m\ge2$. As noted above, including $(p^e,m)=(2,2)$ does no harm.
Let $A_{e,m}$ be the set of lengths derived from $(n_{e,m},p^e)$.
Since $\ord_{n_{e,m}}(p^e)=m\ge2$, Proposition~\ref{prop:fixed}
gives $d(A_{e,m})=0$. Also $A_{e,m}\subseteq n_{e,m}\bbN$, and
\[
 \sum_{e\ge1}\sum_{m\ge2}\frac1{n_{e,m}}
 \le2\sum_{e\ge1}\sum_{m\ge2}p^{-em}
 =2\sum_{e\ge1}\frac{p^{-2e}}{1-p^{-e}}
 \le\frac2{1-p^{-1}}\sum_{e\ge1}p^{-2e}<\infty,
\]
so Lemma~\ref{lem:tails} provides the desired conclusion.

For the ovoid family, let $n_e=(p^e-1)(p^{2e}+1)$, where $e\ge1$,
and let $A_e$ be the set of lengths derived from $(n_e,p^e)$.
Since $\ord_{n_e}(p^e)=4>1$, Proposition~\ref{prop:fixed}
gives $d(A_e)=0$. Also $A_e\subseteq n_e\bbN$, and since $n_e\ge p^{3e}/2$, we have 
\[
 \sum_{e\ge1}\frac1{n_e}
 \le2\sum_{e\ge1}p^{-3e}<\infty.
\]
Therefore, \Le{lem:tails} again provides the desired conclusion.
\epf


\subsection{\label{sec:es-density}Density zero for lengths in the equally-spaced family}
%
Let $\ESset_q$ be the set of all $n$ satisfying~\eqref{eq:es}. 
As before, including $n=4$ when applicable does no harm. Thus
\[
\ESset_q=\{n\in\bbN:\gcd(n,q)=1,\ 
\exists\,d,s\in\bbN,\ d,s\ge2 \quad [n=ds,\quad
\ord_n(q)=s\ord_d(q)]\}.
\]
We first prove density zero when the prime $r$ in Lemma~\ref{lem:witness} below
and its exponent in $n$ are fixed. We then use Lemma~\ref{lem:tails}
to obtain the result for all of $\ESset_q$.

\ble{lem:witness}
An integer $n$ coprime to $q$ belongs to $\ESset_q$ if and only if
there is a prime $r\mid n$ such that
\begin{equation}\label{eq:witness}
 \ord_n(q)=r\ord_{n/r}(q).
\end{equation}
Every such prime satisfies $r^2\mid n$.
\ele
\bpf
This follows from \cite[Lemma 4.3 and the proof of Proposition 4.4]{FHLX}.
For completeness, we give a short proof.

First suppose $n\in \ESset_q$. Then there are $d,s\geq2$ with $n=ds$  and $\ord_n(q)=s\ord_d(q)$. 
Choose a prime divisor $r$ of $s$. By Lemma~\ref{lem:order-bound} 
(applied twice), we obtain
\[
\ord_n(q)\le r\ord_{n/r}(q)
\le r\frac{s}{r}\ord_d(q)=\ord_n(q).
\]
Thus $\ord_n(q)=r\ord_{n/r}(q)$. So in the definition of $\ESset_q$, we may assume that $s$ is prime.

If $r$ divides $n$ only once, then $\gcd(r,n/r)=1$, hence $\ord_n(q)=\lcm(\ord_r(q),\ord_{n/r}(q))$.
Then
\[
 r=\frac{\ord_n(q)}{\ord_{n/r}(q)}=
 \frac{\lcm(\ord_r(q),\ord_{n/r}(q))}{\ord_{n/r}(q)}
 =\frac{\ord_r(q)}{\gcd(\ord_r(q),\ord_{n/r}(q))}
\]
and $\ord_r(q)$ divides $r-1$, so this gives a contradiction. Thus~\eqref{eq:witness}
implies $r^2\mid n$.
Conversely, if a prime $r\mid n$ satisfies~\eqref{eq:witness}, then $r^2\mid n$ by the preceding argument, so $n/r\ge2$. Taking $s=r$ and $d=n/r$ shows that  $n\in\ESset_q$.
\epf

For each prime $r\nmid q$ (equivalently, $r\neq p$), let
\[
 \ESset_{q,r}=\{n\in\ESset_q:r\mid n,\ \ord_n(q)=r\ord_{n/r}(q)\}.
\]
Lemma~\ref{lem:witness} gives
$\ESset_q=\bigcup_{r\nmid q}\ESset_{q,r}$, where the union is over primes.
Note that a length may belong to more than one $\ESset_{q,r}$.
Write $\nu_r(n)$ for the exponent of $r$ in $n$.

\ble{lem:fixed-witness}
For each prime $r\nmid q$ (equivalently, $r\neq p$) and each integer $a\ge2$, the set of
$n$ in~$\ESset_{q,r}$ with $\nu_r(n)=a$ has density zero.
\ele
\bpf
Write $n=r^a v$, where $r\nmid v$, and let
$k=\ord_{r^a}(q)$, $k'=\ord_{r^{a-1}}(q)$, and $t=\ord_v(q)$.
Since $\gcd(r,v)=1$, we have $\ord_n(q)=\lcm(k,t)$ and $\ord_{n/r}(q)=\lcm(k',t)$, and since $n\in\ESset_{q,r}$, we obtain $\lcm(k,t)=r\lcm(k',t)$. By~\Le{lem:order-bound}, we have $k'\mid k$, so from the preceding, we conclude that $k\ne k'$ and hence $k>1$.

Next, we claim that $v$ has no prime divisor in $\PrimeSet(q,k)$.
Indeed, if a prime $s\in\PrimeSet(q,k)$ divided $v$, then
\[k'\mid k\mid\ord_s(q)\mid t,\]
which would give $\lcm(k,t)=t=\lcm(k',t)$, contradicting the preceding $\lcm(k,t)=r\lcm(k',t)$.

Here $k\geq2$ depends only on $q$, $r$, and $a$, so \Tm{thm:wiertelak} and \Le{lem:sieve} give density zero for the possible $v$. Since $r^a$ is fixed, the corresponding lengths $n=r^a v$ also have density zero.
\epf

\btm{thm:es-zero}
For every fixed prime power $q$, the set $\ESset_q$ has natural density zero.
\etm
\bpf
By Lemma~\ref{lem:witness},
\[
 \ESset_q=\bigcup_{\substack{r\text{ prime}\\r\nmid q}}\ \bigcup_{a\ge2}A_{r,a},
 \qquad A_{r,a}=\{n\in\ESset_{q,r}:\nu_r(n)=a\}.
\]
Each $A_{r,a}$ has density zero by Lemma~\ref{lem:fixed-witness}
and is contained in $r^a\bbN$. Moreover, by~\eqref{eq:primepower-sum},
\[
 \sum_{\substack{r\text{ prime}\\r\nmid q}}\sum_{a\ge2}\frac1{r^a}\le1.
\]
The claim now follows from~\Le{lem:tails}.
\epf

\subsection{Proof of Theorem~\ref{thm:main}}\label{sec:proof}
Fix $p$. By Theorem~\ref{thm:classification} and
Proposition~\ref{prop:tensor}, every non-standard pair $(n,p)$ is derived
from one of the five initial families in characteristic $p$.
If the initial pair is equally spaced, Proposition~\ref{prop:closed}
shows that $(n,p)$ is equally spaced; its lengths have density zero
by Theorem~\ref{thm:es-zero}. 
The other four families contribute density zero by
Corollary~\ref{prop:golay} and
Propositions~\ref{prop:repetition} and~\ref{prop:primitive-ovoid}.
Hence~$\NSset_p$ has
density zero. 
Corollary~\ref{cor:prime-reduction} gives
$\bigcup_{r\ge1}\NSset_{p^r}=\NSset_p$ and proves the
asserted simultaneous statement.
Finally,
\[
 \#\{n\le X:\gcd(n,p)=1\}=(1-1/p)X+O(1),
\]
so division by this count gives relative density zero for the
exceptional lengths, and relative density one for their complement.
\hfill$\square$

\section{Conclusions}\label{sec:conclusions}
The Berger--Charpin conjecture predicts that cyclic codes
generally have only the permutation automorphisms arising
from cyclic shifts and Frobenius maps. 
We have established this expectation in a precise asymptotic
sense for non-degenerate irreducible cyclic codes 
(including degenerate irreducible cyclic codes would instead
give a set of non-standard lengths of relative density 1 among lengths coprime to the characteristic).
Concretely, we show that for each fixed prime~$p$,
the set of lengths admitting an irreducible non-degenerate cyclic
code with additional permutation automorphisms over some finite
field of characteristic $p$ has natural density zero. Thus, for each fixed prime~$p$ and almost every length coprime
to $p$, every non-degenerate irreducible cyclic code of that
length over any finite field of characteristic $p$ has only
the standard permutation automorphisms.

The proof combines the classification of these codes with
number-theoretic results on the multiplicative orders of integers
modulo primes. 
We describe the lengths obtainable from the five families in the
classification and show that, for each fixed characteristic,
each family together with all codes obtained from it by the
constructions contributes a set of lengths of density 0.



\section*{Acknowledgments}
This research was supported by the National Key Research and Development
Program of China (Grant No. 2025YFA1017700). Tao Feng was supported by
the National Natural Science Foundation of China (Grant No. 12225110).
Henk D.L. Hollmann was supported by the Estonian Research Council grant
PRG2531. Weicong Li was supported by the National Natural Science Foundation
of China (Grant No. 12301422). Qing Xiang was supported by the National
Natural Science Foundation of China (Grant No. 12131011).

The authors acknowledge the provision of access to OpenAI models through the ChatGPT for Academic Researchers program. They used ChatGPT (Astra model), under their guidance, to develop and simplify mathematical arguments, assist in proofreading, and improve the English writing. All proofs in the manuscript were 
written by the authors, who take full responsibility for their correctness, and for any remaining errors.

\bigskip
\begingroup\small
\noindent Tao Feng, School of Mathematical Sciences, Zhejiang University,
Hangzhou, China.\\
Email address: \texttt{tfeng@zju.edu.cn}

\medskip\noindent Henk D.L. Hollmann, Institute of Computer Science,
University of Tartu, 51009 Tartu, Estonia.\\
Email address: \texttt{henk.d.l.hollmann@ut.ee}

\medskip\noindent Weicong Li, Department of Mathematics, School of Sciences,
Great Bay University, Dongguan, China.\\
Email address: \texttt{liweicong@gbu.edu.cn}

\medskip\noindent Qing Xiang, Department of Mathematics and Shenzhen
International Center for Mathematics, Southern University of Science
and Technology, Shenzhen, China.\\
Email address: \texttt{xiangq@sustech.edu.cn}
\endgroup
\end{document}